# THE FLOWING SYSTEM GASDYNAMICS
## Part 1: On static head in the pipe flowing element

S.L. Arsenjev, I.B. Lozovitski[1], Y.P.Sirik

*Physical-Technical Group*
*Dobroljubova street 2, 29, Pavlograd, Dnepropetrovsk region, 51400, Ukraine*

The solution of problem on distribution of static head along the pipe flowing element is submitted. The solution is reached on the basis of consideration of contact interaction of gas and liquid stream with the wall of the pipe flowing element. The expression for distribution of static head along the pipe flowing element is obtained on the basis of usage of three fundamental laws in fluid dynamics: Torricelli formula, Weissbach-Darcy formula and Bernoulli equation. The general solution is obtained for a gas stream. The special case of the obtained solution is retrieved for liquid stream.

## Nomenclature

| | |
|---|---|
| $\gamma$ | weight density of fluid medium |
| $g$ | acceleration of gravity |
| $h_i$ | height of current point at free fall |
| $H$ | general height of fall |
| $\lambda$ | coefficient of hydraulic friction |
| $l$ | length of stream from the inlet section of flowing element up to current point |
| $L$ | general length of flowing element |
| $\overline{L}$ | general caliber length of flowing element, L/D |
| $D$ | internal diameter of flowing element |
| $\zeta_{in}$ | coefficient of local hydraulic resistance for inlet into flowing element |
| $\zeta_{ex}$ | coefficient of local hydraulic resistance for outlet from flowing element |
| $p_0, p_h$ | quantities of pressure on inlet and outlet of flowing element accordingly |
| $p_{st}$ | static head in stream |
| $V_{fw}$ | stream velocity determined by weight flow |

## 1 Introduction

The creating of physically substantial bases of the fluid medium motion theory envisions overcoming the set of problems and solution of series of attendant questions. The problem of determination of distribution of the energy potential along the pipe flowing element or system is one of fundamental in this area. The essence of this problem lies in the fact to define an unitized mathematical expression, which will physically adequately and technical exactly determine the distribution of energy potential applied to the flowing element or system and it will be equally fairly for liquid and gas streams. In applied sense it means to determine of the law of static head distribution along the pipe flowing element or system.

## 2 Approach

For the solution of the problem the authors utilized: the formula for free fall setting by proceedings of Torricelli - Galilei (1643) and Borda - Du Buat (1766), the formulae for determination of friction losses at motion of liquid stream in pipe of Weissbach-Darcy (1857) and local losses of Weissbach (1865) and also the equation of conservation of energy (density of energy is more exact) for stream in pipe by Bernoulli principle (1738) in the form that conform to the end of XIX century. The sufficiency of these dependences for reaching an object is determined by their stated below distinctions. Torricelli - Galilei - Borda - Du Buat (TGBD) formula represents the historically first equation of conservation of energy for motion of solid or medium without contact interaction. The path of development of this formula from the moment of an experimental research up to physically correct record was required almost 150 years (Du Buat) and up to usage in equation of conservation of energy was required additional almost 120 years. This formula has passed inspection in general mechanics and then in mechanics of fluid medium and it is one of fundamental in mechanics. Weissbach-Darcy's formula represents the historically first modification of TGBD formula for fluid medium motion with taking into account of contact interaction (with pipe wall). This formula is obtained also on the basis of experimental research and reflects the approach to a viewing of motion indicated by Aristotle (328 B.C.) as the third problem of his «Mechanical problems». The fundamentality of taking into account of contact interaction called simplistically as hydraulic friction is affirmed not only Aristotle's prevision but also that this formula became an inalienable part of equation of conservation of energy for fluid flow in pipe. Weissbach's formula is modification of Weissbach-Darcy's formula and allows to determine the different shape local resistances at fluid medium motion in flowing system.

---

[1] Phone 38 05632 38892, 40596
Electronic address: loz@inbox.ru

The fundamental character of equation of conservation of energy for stream of fluid medium in flowing element or system, apparently, cannot call doubt, and furthermore this equation is constructed on above-mentioned laws.

The principle of conservation of energy at motion without contact interaction and the principle of consumptions of energy for contact interaction at motion is harmonic combined in this equation. The last reflects even more severe sense, connected with self-organizing of stream with minimum of energy consumption that applied to the flowing system.

3 Solution

So, TGBD formula may be written in the form:

$$\frac{V^2}{2g} = H\left(1 - \frac{h}{H}\right) \qquad (1)$$

where $h$ represents simultaneously part of not yet passed path by impinging body and part of residual potential energy and $H$ represents simultaneously all path of falling and available initial value of potential energy. In such aspect this formula allows to determine the quantity of kinetic energy of falling body in dependence on relative part of passed path of falling.

In its turn Weissbach-Darcy's formula in the form:

$$p_{st}(l) \equiv \gamma h(l) = \lambda \overline{L}\left(1 - \frac{l}{L}\right)\frac{\gamma V_{fw}^2}{2g} + p_h \qquad (2)$$

allows to determine the size of static head in liquid stream along the pipe depending on intensity of contact interaction of fluid with wall. The reference point of current length of stream start from inlet section of pipe just as the reference point of current fall height in the TGBD formula start from point of falling.

Taking into account of denotation $\left(1 - \frac{l}{L}\right) = K_l$

Weissbach-Darcy's formula assumes the compact form:

$$p_{st}(l) \equiv \gamma h(l) = \lambda \overline{L} K_l \frac{\lambda V_{fw}^2}{2g} + p_h \qquad (3)$$

The equation of conservation of energy for liquid stream in direct pipe of round section with taking into account of formula (3) and Weissbach's formula becomes:

$$p_o = p_h + \frac{\gamma V_{fw}^2}{2g} + (\lambda \overline{L} K_l + \zeta_{in} + \zeta_{ex})\frac{\gamma V_{fw}^2}{2g} \qquad (4)$$

After determination $V_{fw}^2$ from equation (4) and substitution it in the formula (3) we receive:

$$p_{st}(l) = (p_0 - p_h)\frac{\lambda \overline{L} K_l}{1 + \lambda \overline{L} K_l + \zeta_{in} + \zeta_{ex}} + p_h \qquad (5)$$

The expression (5) determines distribution of static head in stream of fluid medium in pipe.

According to expression (4) for liquid the equality is valid:

$$V_{fw}^2(l) = \frac{2g}{\gamma}(p_0 - p_h) \times$$

$$\times \frac{1}{1 + \lambda \overline{L} K_l + \zeta_{in} + \zeta_{ex}} = const \qquad (6)$$

From equality (6) it is follows necessity to accept $K_l = 1$, executable by $l = 0$, what corresponds to the outlet section of pipe. Therefore for liquid the distribution of static head along pipe has accordingly kind:

$$p_{st}(l) = (p_0 - p_h)\frac{\lambda \overline{L} K_l}{1 + \lambda \overline{L} + \zeta_{in} + \zeta_{ex}} + p_h \qquad (7)$$

and testifies about linear change of static head along the stream length in pipe according to change of the fraction numerator in its right member.

In contrast to this, expression (5) possess the generality and is applied to stream of gas medium in pipe. Presence of a complex varying along the pipe length $0 \leq K_l \leq 1$ both in numerator and in denominator of fraction in a right member of expression (5) specifies the nonlinear character of static head change in gas stream in pipe. At the same time the diminution of quantity of static head in gas stream in direction of outlet from pipe happens more and more intensively. Physically it means increasing of velocity of gas stream to outlet from pipe with constant values of cross section and coefficient of hydraulic friction. Such differences of physical characteristics of gas from liquid as a major elastic (reversible) compressibility and a major kinematical viscosity in combination with small heat capacity stipulate the consecutive transformation of friction work to the heat, which one, in turn, stipulates the consecutive increase of gas stream velocity along the pipe of uniform cross section. In distribution of static head along the pipe for liquid and gas, the common is that the initial and final values of static head can be the same for these two mediums. The difference is, that at the mentioned identity, the curve of static head of gas stream between points $p_{st}(0) = p_0$ and $p_{st}(L) = p_h$ is situated above a straight line of static head of liquid stream.

In problem about free fall in gravity field, the velocity of body is increased on the way of the fall under stationary value of gravity acceleration. In problem about motion of gas stream in pipe the flow velocity is increased under stationary value of intensity of contact interaction of gas stream with surface. Such acceleration of gas stream motion is the frictional self-acceleration.

## 4 Discussion

Lately, the analogous object was pursued by Duxbury in his work [1]. The equation derived by this author contains the link of pressure drop along the pipe with the mass flow rate of fluid. The link is not immediate because it requires to introduce the discharge coefficient as the additional empirical factor defined experimentally. Furthermore, the link of pressure drop with flow rate is complicated by the necessity of determination of density of fluid at the beginning and end of stream as well as an average quantity in conditions of essential non-linear nature of the change of the pressure drop along the pipe. The mentioned imperfections is a consequence of the physically inferior approach based on balance of the forces but not the equation of conservation of energy. Therefore equation for determination of pressure drop along the pipe derived in [1] as many other attempts before him no possess the attributes of the static head law along the pipe flowing element.

## 5 Congruence of results

The obtained expressions (5) and (7) are used as basis for working out of mathematical algorithm and then VeriGas program for computing of the state and motion parameters of gas stream in the diverse flowing element and its systems. The computational results are tested by means of comparison with experimental data of a great many of sources of the specialized literature.

In particular, it was determined, that overstated experimental curve of static head approximately on 5% in contrast to our computations, inserted in the book [2], is explained that the pipe termed in the book text as a smooth really had 19 static head measurement orifices. Thus, the comparison has given result not for the benefit of experiment. There is a lot of such examples in papers and monographies [3,4,5].

## 6 Final remark

The expressions shown in this paper have a key nature, therefore their deduction is submitted in one-dimensional stationary statement. At the same time, results of computational experiments that repeatedly realized does not yield to experimental results on regularity and precision.